# Voronoi Construction for Improving Numerical Calculations of Particles

Wilfried Wunderlich,
Nagoya Institute of Technology, Dept. of Environmental Technology,
Gokiso Showa-ku, 466-8555 Nagoya, e-mail: wunder@system.nitech.ac.jp

**Abstract**
Calculations on atomistic scale are necessary for understanding of physical phenomena occurring during advanced processing of liquids, slurries, and nano-ceramics composite materials. This paper describes some new ideas for using the Voronoi construction for calculations of amorphous materials or liquids in comparison to the regular arrangement of atoms in crystals. This approach for treating numerical calculations can be used for any particle simulation method, like Molecular Dynamics or Ab-initio calculations in real or reciprocal space. In the discussion the interpretation of electron dispersion relations for use in solid state physics is also summarized.

## 1. Introduction

Calculations of irregular arrangement of particles, like in liquids or amorphous materials, require special methods for numerical treatment of properties. The Voronoi construction [1] can improve this analysis of the data. The method described here can be applied for example for the challenging processing of nano-ceramics composite materials. Compared to conventional materials the processing techniques for nano-particles are more complicated, since they are less stable and tend to collapse to larger agglomerates. In a previous overview [2] examples of ab-inito calculations of ceramic nano-particles and interfaces were shown. This calculation method shown in this paper is useful for particles in an irregular pattern no matter, on which length scale they are considered. Especially for MD or ab-initio calculations the Voronoi analysis can save a lot of calculation time. Furthermore, when applying it in reciprocal space, the theory of electronic calculations and band structure calculations can be improved, for reaching the goal, to develop new materials or improve materials processing by meaningful simulations. This paper gives in the following some outlines, since this subject covers a huge area and is still developing very fast.

The paper is divided into three parts: first, the needs for simulation of advanced materials like simulations of species in aqueous solutions, the physical properties, or and nano-particles, the geometry of simulations, the calculation methods, and the physical meaning of simulations are outlined.

## 2. Needs for simulation
### 2.1. Material development

The needs for ab-initio- and simulations on atomic scale (fig. 1) can be divided in screening, microstructure improvement, analyzing new physical effects and process modeling. The screening focuses on the search for new material compositions, composite materials, or special geometries like in gradient materials. Expensive experiments like e.g. specimens containing rare earth or the analysis of hydrogen, require simulations, which are relatively cheap. The microstructure of devices can be improved with the aid of simulations, to find optimal particle coating or thin film microstructures. The influence of segregation, grain boundary or interfaces structures [3], vacancies, or interstitials can be calculated in the computer more easily compared to the time-consuming experiments. New physical effects have been found by ab-initio calculations, like nano-size effects in ceramic particles, electron excitation effects, electron coupling etc., and also their understanding can be improved by simulation, in order to find better thermoelectric materials, electro-ceramics, photovoltaic materials, or precursor materials. For research in ceramics engineering, calculations can support the understanding of occurring phenomena in chemistry, physics or engineering, and problems are much easier to solve, when the reason can be found. Finally, for ceramics processing the understanding and data about the chemical reactions, like reaction probability, reaction path or sequence of the reactants or reaction velocities are essential. When using polymers the understanding of bonding to the inorganic surface is important. When applying the material for photo catalysis, photochemistry, or photovoltaic cells, the understanding of the photon excitement by simulation is essential, as described in the next chapter. Finally, for the understanding of phenomena occurring during processing, like slip-casting, the adhesion of particles etc. simulations are necessary as described in chapter 2.2.

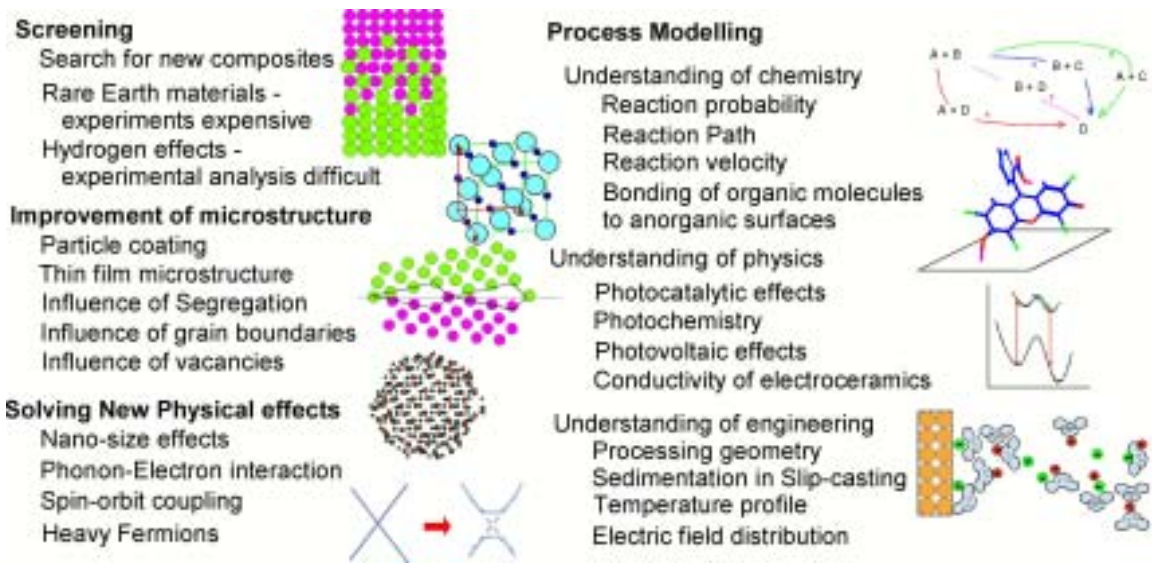

Fig. 1 Needs for atomistic and ab-initio- calculations in advanced materials processing

## 2.2. Calculation of nano-particle interaction

Optimization of nano particle processing requires a balance of attractive and repulsive forces between the particles. For the aqueous solution used for ceramics processing several different method of stabilization have been developed [4-7]. The most effective repulsion is performed by Coulomb forces caused by polarization of the particle surface either in the acid or base region left or right of the iso-electric point. This electrostatic repulsion can be performed by ions or charged polymer chains or both. On atomic scale the charge distribution on particle surfaces has been explained by the DLVO theory [4-5]. Figure 2 shows the case, that cations have a higher absorption ratio to the ceramic surface, so that a negative charge at the surface is formed. Hence, the probability is high, that in the next layer of molecules between the inner and outer Helmhotz layer (IHP, OHP) positive charged particles are in the majority. In the subsequent stern layer the potential reaches the level of the aqueous solution. In the case, that anions have a higher absorption rate, the IHP and OHP have opposite charge. For quantitative understanding of these phenomena simulation is necessary.

## 3. Modeling of particle interactions
### 3.1. Molecular Dynamics

In molecular dynamics (MD)-calculations the dynamic behavior of atoms is modeled by assuming a suitable interaction potential or force field between the atoms. (In similar way the discrete element method calculates the interaction between mesoscopic particles [8-9]). A MD- program called Yasp [10-11] is available, in order to simulate the dynamics and properties of polymers or other molecules in liquids [10-12], but for its application the search of suitable MD- potentials is necessary, which can be difficult, because the potential parameters can depend on the liquid environment. For crystals, where the environment is almost unchanged, suitable parameter sets for the Buckingham/Morse potential for $Al_2O_3$, MgO and $MgAl_2O_4$ can be found by fitting the parameters to macroscopic properties, like lattice constants, and thermal expansion coefficient of single crystal experiments. A suitable software for MD-simulations is Moldy [13], which calculates the forces in reciprocal space and is especially suitable for ionic and covalent bonded materials.

By MD the interface and surface energies can be calculated with high accuracy [14] and it would be interesting to compare the influence of the geometry. The typical size for supercells in affordable MD calculations exceeds already 20000 atoms and hence, also mesoscopic effects can be calculated, like misfit dislocations, interaction of dislocations, particle interactions etc. However, the MD results strongly depend on the reliability of the potentials and their search can take a long time. Furthermore, it cannot be excluded, that the inter-atomic potentials will change at defects. That is the reason, why ab-inito methods are becoming more and more popular, although affordable calculations are restricted to about 100 atoms. They are described in chapter 3.3, but before that, a mathematical construction useful for both MD or ab-initio is considered.

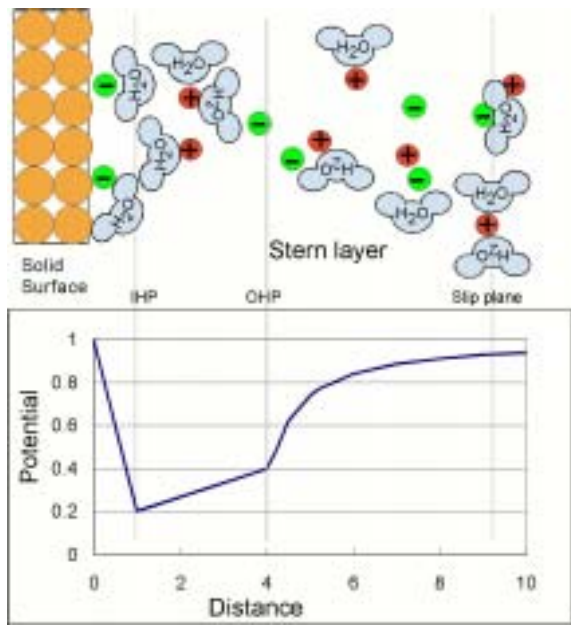

Fig.2 Detail on atomistic scale of the electric potential at the particle surface

**3.2. Modeling of aqueous solutions by the Voronoi method**

The following mathematical method is still under development and might be in future a promising tool to accelerate any particle calculation method, especially for treating particles or atoms in an amorphous environment like liquid. The Voronoi construction is a numeric mathematical method for fast and effective characterizing of many microstructural properties [1], like the empty space between particles, the nearest neighbor distances, open channels and their distribution, etc. The particles can be conventional ceramic particles on macroscopic scale, nano-particles or atoms and can have the same radius, a bimodal distribution, or a broad distribution of radii.

The construction can be performed in two ways (fig. 3), in the following called Voronoi- and Wigner-Seitz-method. In the first method, lines are drawn from the center of each particle to their nearest neighbors. These lines form polygons (fig. 3a) or, in three-dimensions, polyhedrons, which are equivalent to the Bernal polyhedrons describing e.g. the atomic arrangement in amorphous materials or liquids. The next step is to find the largest sphere inside these polyhedrons and their center defines the point set of the Voronoi diagram, which lead to further analysis of the microstructure, like the broadest path, largest empty space etc. [1]. The second method (fig. 3b) is the Wigner-Seitz-cell construction, in which the shortest distance between the particles is divided into its half and the spatial distribution of the endpoints of all of these vectors leads to polyhedrons with their centers at the particle center. This method is used for the construction of the Brillion Zone in reciprocal space. In the Wigner-Seitz construction the polyhedron is a measure for the environment affecting the particle, while in the Voronoi method the sphere describes the empty space between particles and it depends on the actual physical problem, which method is the more suitable one.

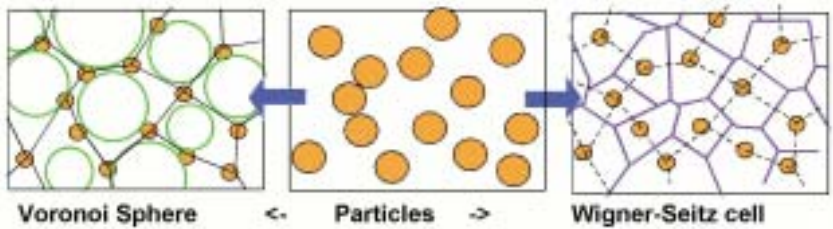

Fig. 3 Construction of Voroni spheres and Wigner-Seitz cells

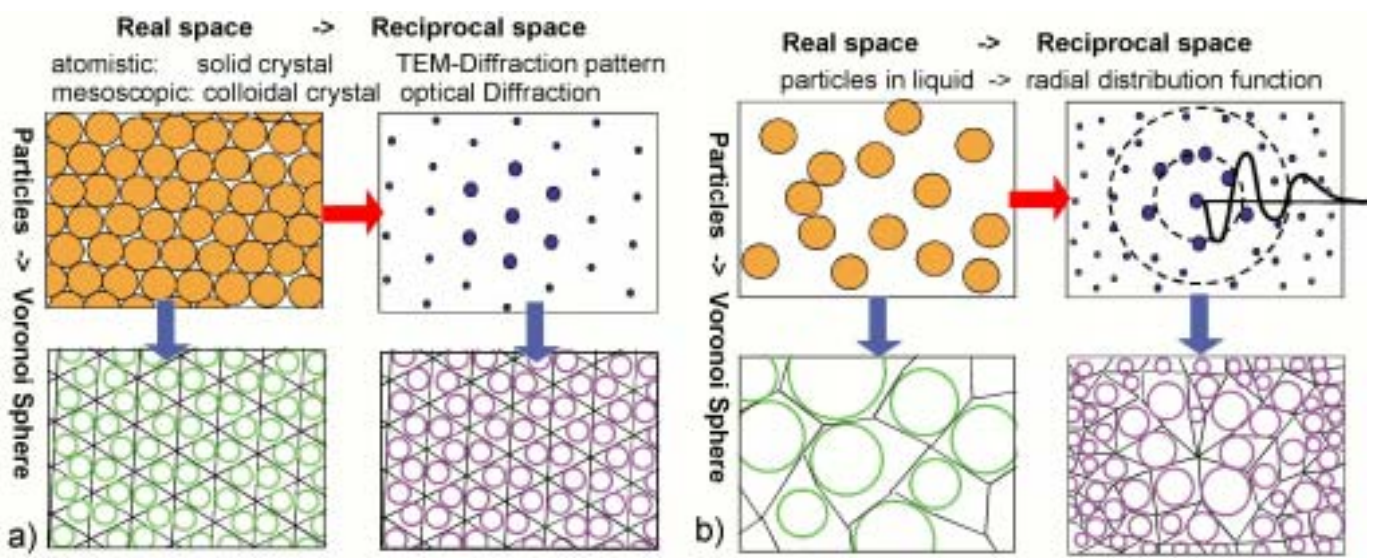

Fig. 4 Schematic drawing of the transformation from real space to reciprocal space, and the Voronoi construction,
a) for atoms in a crystal or for nano-particles on mesosopic scale,
b) for diluted atoms or particles in aqueous solutions

The regular arrangement of equi-radii particles, e.g. atoms in closed packed crystals or particles in a colloidal crystal (fig. 4a), leads to a regular pattern of the Voronoi spheres. Also, if diffraction in reciprocal space on these particles is considered, like electron diffraction at atoms or optical diffraction on colloidal particles, a regular point pattern with high symmetry is obtained. The Voronoi diagram of this pattern is the same as in case of hexagonal symmetry. In the case of *bcc* it transforms to *fcc* pattern and vice versa. Due to the high symmetry also equi-radii spheres are obtained.

The situation changes, if we consider now particles in a liquid (fig. 4b). Even when the particles have the same radii, the Voronoi spheres have random sizes. The diffraction pattern in reciprocal ($k$-) space, e.g. obtained in the computer by fast Fourier transformation (FFT), shows a random distribution of diffraction spots in the outer regions, but a certain accumulation of spots in a first and second ring near the center spot as described by the radial distribution function (RDF, inlet in Fig. 4b). The reason for the RDF-shape with a clear first maximum, and dissipation for larger $k$-distances is the balance between adhesion and repulsive forces between the particles. The Voronoi construction of this diffraction pattern leads to polyhedrons with mostly elongated shapes containing large circles in their center. Beyond the rim of the first maximum in the RDF the spheres become smaller and smaller, and the polyhedron sphere-like. It would be a challenge to find out, whether there is a simple transformation between the Voronoi sphere in real and reciprocal space and also how the corresponding Wigner-Seitz cells will transform.

The characterization of the microstructure by such mathematical methods is necessary to study the physical properties. The polyhedron in reciprocal space describes the scattering behavior of waves, and the schematic drawing (fig. 4b) shows, that only the large polyhedrons near the center are significant for the physical properties of the material. Using the Voronoi construction in reciprocal space physical effect for particles in liquids can be easily analyzed, like the Bragg law of scattering electro magnetic waves, the excitation of photons, or calculating the conductivity of electrolytes or amorphous materials. This Voronoi construction becomes depended on the position, when the particle distribution is inhomogeneous, like at sedimentation or in the vicinity of electrodes. The application of this method will be progressing, when it could be adapted to common MD- or *ab-initio-* software packages.

### 3.3. *Ab-initio* calculations

The most sophisticated and powerful calculation method for material science is the first-principle or *ab-initio*-method, which is based on the density functional theory (DFT) [15] and needs no empirical parameters. The many-body problem of calculating electrons in the Coulomb potential of the atom cores is simplified by assuming a single electron affected by a pseudo-potential, including the contributions of the atomic core, the core electrons, and other electron wave functions. The theory was developed by starting from the three extreme cases, nearly free electrons, tight binding and electron levels at covalent bonds. The charge density is calculated self-consistently in the computer by an appropriate software program. Such a program is Vasp [15-17], which is well accepted in the community and widely used for its reliability. It includes many facilities, like e.g. calculating the forces between atoms, which allows the atom migration into relaxed positions and make dynamic studies possible. Vasp has the advantage, that it calculates absolute energy values, which only depend on the element and its input pseudo-potential [17]. Hence, the output total energy can be directly used for comparing different input structures and this is one of the ways to use electronic calculation as a vehicle for material science application. In more sophisticated applications the density of states and the electron dispersion curve are calculated and interpreted as explained in the last chapter of this paper.

Many material science problems have successfully been calculated by Vasp: Adsorption of molecules on surfaces [18-20], transport properties in liquids [21], colors of intermetallics [22], etc. Also the electric conductivity of amorphous materials [23] has been calculated from first principles by splitting the band gap into gaps with different depths and widths. A Greens-function concept correlated this distribution with an average pseudo gap, and using the Kubo-Greenwood theory the calculated conductivity is in good agreement with the experimentally measured one. However, unrealistic scattered waves appeared due to limited supercell, which had to be distinguished by artificial virtual sources or filtered out in reciprocal space and perhaps the promising Voronoi approach described in the previous chapter can simplify these calculations.

### 3.4. Interpretation of Band structure calculations

By *ab-initio* calculations band structure diagrams are obtained, which need a proper interpretation in order to understand and optimize optical and electrical properties of materials. Figure 5 shows typical features occurring in such energy dispersion curves in the reciprocal space ($k$-space) as deduced from textbooks [24-26]. The Fermi-Energy is marked as dashed line and the band width or overlap between bands (Fig. 5a) can be estimated directly from the difference between the energy states and characterizes, whether a material is an insulator or semiconductor because of an existing band gap, a metal or semimetal because of band overlap or half-filled band. For the conductivity the number of charge carriers are responsible. Bands with a local minimum collect electrons in so-called electron pockets, which can be below or in the case of thermal excitement slightly above the Fermi energy. A hole pocket is a local maximum slightly above the Fermi energy. Doping of semiconductors changes the level of the Fermi-energy relative to the band structure, as occurring e.g. in PbTe.

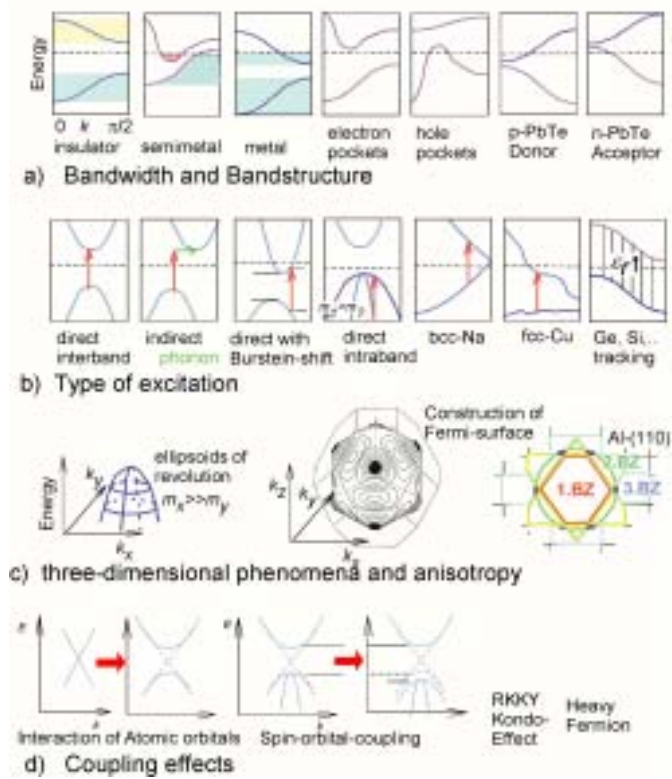

Fig. 5 Information obtained from electronic band structure calculations

For optical properties the type of excitement (fig. 5b) is important. The direct interband transitions have a higher probability than the indirect transitions, because they require phonon scattering. When the extrema of the two bands are not at the same $k$-vector, a so-called Burstein shift occurs: The excitement needs a higher energy than from the band gap expected. Bands with a smaller curvature in the $E(k)$ diagram lead to higher effective carrier masses, which affect their mobility and hence the conductivity. In some cases the bands with high or low effective masses are overlapping, so that intraband transitions are possible. Special features are observed in Na with bcc-structure, which is almost like a free electron and hence a photon excitement with visible light (1-2eV) is possible. In fcc-Cu optical transition between a lower band and electron pockets is occurring, and this is one of the reason for its reddish color. In some semiconductors like Si or Ge two bands are almost parallel over a wide range of $k$-vectors and this so-called tracking leads to a high dielectric constant.

Interpretation of the electron dispersion curves in three-dimensions (fig. 5c), either as $E(k_x,k_y,k_z)$ or as iso-energy-lines in the $k$-space, leads to the constructing the Fermi-surface, which can be very complicated for multi-valent metals (Cu, Al, etc.) and explains anisotropy effects. Finally, coupling effects between electrons occur (fig. 5d), which usually are not included in the conventional band-structure calculations based on the single-electron approximation. A common method is, to apply a post-calculation treatment on the band-structure, so that details like the interaction of atomic orbitals or the enlargement of the band-gap due to spin-orbital coupling appear in the final electron dispersion curve. However, strong coupling effects, like magnetic spin interaction (RKKY-effect), Kondo-effect, or heavy fermions are multi-electron effects in materials with d- and f-electrons are still lacking in understanding and numerical modeling.

## 4. Summary

This paper summarizes the needs of simulation in material science, and showed some new ideas for improving the numerical treatment of particles in liquids by the Voronoi method. This method is useful for analyzing the interaction of nano-particle and in the slurry. The prevention of their agglomeration is the main goal for nano-ceramics processing in future, and it is a challenge to find this by simulation. Another aim is, to study the electron dispersion curves of amorphous or liquid materials in reciprocal space by ab-initio software. For both, the MD and the ab-initio method the Voronoi construction could be a useful tool. Also the present density functional theory, which is the base for the present *ab-initio* software, is limited to the one electron concept and hence, strong correlations are neglected, which would be necessary to improve for example thermoelectric materials. It is just a matter of time, computer development and the number of simulation-software users, that such problems will be overcome in future.